\documentclass[12pt]{article}
\usepackage{graphicx}        
\usepackage{amsfonts}
\def\ll{\label}

\def\r1{(\ref{$1})}

\def\ba{\begin{array}{c}}

\def\ea{\end{array}}

\def\l{\left}
\def\l({\left(}
\def\r){\right)}
\def\r{\right}

\def\la{\lambda}

 \def\be{\begin{equation}}
\def\bc{\begin{center}}
\def\ec{\end{center}}
\def\bit{\begin{itemize}}
\def\eit{\end{itemize}}
\def\ee{\end{equation}}
\def\ed{\end{document}}
\def\bea{\begin{eqnarray}}
\def\eea{\end{eqnarray}}
\def\efr{\end{flushright}}

\begin{document}
\title{{
 Nonholonomic   deformation of  KdV and mKdV
equations and their symmetries, hierarchies and integrability
 }}
\author{ Anjan Kundu \footnote{anjan.kundu@saha.ac.in}\\
 Theory Group \& CAMCS,
Saha Institute of Nuclear Physics\\
 Calcutta, INDIA\\
R. Sahadevan \footnote{ramajayamsaha@yahoo.co.in} and L. Nalinidevi \footnote{lnalinidevi@yahoo.co.in}\\
Ramanujan Institute for Advanced Study in Mathematics,\\
 University of Madras , Chennai- 600 005,Tamilnadu, INDIA} \maketitle
\begin{abstract}
Recent concept of  integrable nonholonomic deformation found
 for  the KdV equation is
 extended    to the mKdV equation and  generalized
to the AKNS system.  For the deformed mKdV equation we  find a matrix
Lax pair, a novel two-fold integrable hierarchy
 and  exact N-soliton solutions
exhibiting unusual accelerating motion. We show that both the deformed KdV and mKdV  systems
possess infinitely many generalized symmetries,  conserved quantities  and a recursion
operator.
 \end{abstract}

\noindent {\it Short title}: Integrable nonholonomic deformations of KdV and
mKdV
\\ \noindent {PACS:} 02.30.lk,
02.30.jr,
 05.45.Yv,
11.10.Lm,
\\ {\it Key Words} Integrable nonholonomic deformation, deformed  mKdV,  KdV and AKNS
systems, Lax pair,
accelerating N-soliton, two-fold integrable hierarchy, generalized symmetry, conservation law,
recursion operator.
\section{Introduction}

Recently discovered  integrable $6$-th order Korteweg de Vries
(KdV) equation \cite{karasu} was shown to represent a nonholonomic
deformation
 (NHD)  of the well known KdV equation preserving its integrability
and exhibiting an integrable hierarchy
  \cite{kuper08}.
In a subsequent development a matrix Lax pair, the  N-soliton solution
through inverse scattering transform (IST) method and
 an intriguing two-fold integrable hierarchy
were found    for
 this particular system by one of the authors \cite{kundu082}. The NHD of the
KdV (dKdV) was shown recently to be a certain form of  self-consistent source equation
(SCSE) allowing particular exact solutions \cite{Yao08}. 

The NHD for such field
theoretical models is given by a constraint in the form of nonlinear  differential
equation involving only $x$-derivatives on a single perturbing function,
which is
deforming the original integrable equation. 
This type of  integrable  deformation  is a relatively  recent discovery,
 which allows also an integrable hierarchy of higher
order deformations.   
On the other hand the construction of SCSE  is a well known concept
\cite{melnikov}, which is represented usually by a coupled system consisting  of
the  original
integrable equation with an additional term made from the eigenfunctions and 2N eigenvalue equations
 of the Lax operator with explicit
dependence on N discrete eigenvalues. In general 
the  eigenfunctions and eigenvalues are complex and  one can also add
another 2N number of evolution equations for the eigenfunctions. 
In a recent development  of such SCSE its   Rosochatius deformation is
proposed \cite{Yao08}. At certain limit or at particular cases the contact
between the lowest order NHD and the SCSE can be found, though the
integrable hierarchy of  higher order deformations seems to be possible only
for the NHD.   

  In a slightly
generalized form this deformed KdV equation can be given as
\bea u_t-u_{xxx}-6uu_x&=& g_x(t,x),\ll{dkdva} \\
g_{xxx}+4ug_x+2u_x(g+c(t))&=&0,\ll{dkdvb} \eea \noindent where subscript denote partial derivatives. Though
the time-dependent arbitrary function $c(t) $ can be absorbed in
 equation (\ref{dkdvb}) by redefining perturbing function $g(x,t) $, we keep
it in the explicit form for our convenience. Recall that, the
KdV and the modified KdV (mKdV) are intimately related partner
systems \cite{solit1}. Therefore it would be natural to expect that
the concept of NHD found for the KdV should also be extendible    to
the mKdV equation. We show here that this expectation is indeed
true by constructing explicitly a novel integrable NHD of the mKdV
(dmKdV) equation, which can  yield    a new integrable  4th order
potential mKdV equation.   We discover for this  integrable deformed mKdV
equation with a nonholonomic constraint a   matrix Lax pair,
 exact N-soliton solutions  and a novel
two-fold integrable hierarchy, similar
 to the result of the dKdV \cite{kundu082}. The solitons, found for both the basic
field and the perturbing function of the dmKdV, show
unusual accelerating (or decelerating) motion. We show that both the deformed KdV and mKdV  systems
possess infinitely many generalized symmetries and conserved quantities,  and a recursion
operator which have been studied until now only for their undeformed counterparts.
The Lie symmetry analysis we perform here for the deformed equations
leads also to explicit construction for the hierarchies of the
 generalized symmetries for both the basic field and the
perturbing  functions. Our analysis of the continuity equation for the dKdV
and dmKdV reveals that the conserved densities for these deformed systems
remain the same as in their original undeformed cases, whereas the current
densities (fluxes) explicitly contain the deforming functions.
This shows the intriguing fact that the nonholonomic
deformations can appear only at the equation level, while the conserved
integrals of motion remain the same under NHD.

Finally we unify   dKdV and dmKdV to  discover an
integrable  nonholonomic deformation for the
 more    general
 AKNS \cite{solit1} system.

The plan of the paper is as follows. In section $2$ the
new NHD of the mKdV equation and the related integrability
structures such as the matrix Lax pair  and the exact
 $N$-soliton
solution are   presented.  Tt is also shown that the dmKdV admits
infinitely many higher order or generalized symmetries,  conserved
quantities and a
recursion operator. In
section $3$ a similar analysis has been carried out for the
dKdV equation. Section $4$
 generalizes  the nonholonomic deformation
to the AKNS system, revealing 
 a novel two-fold integrable hierarchy for all its members  and consequently for the dmKdV
and the dKdV  systems. In section $5$ we give a brief summary of our
results and the concluding remarks.

\section{Nonholonomic deformation of the mKdV equation
}

Integrable equations with NHD should be driven by
 an additional perturbative or deforming  function, which in turn would be   subjected to
 a differential constraint of nonholonomic nature.
 Therefore, analogous  to the deformed KdV
equation (\ref{dkdva}-\ref{dkdvb}) we propose a deformed  modified
KdV equation with nonholonomic constraint as
 \bea
& &v_t-v_{xxx}-6v^2v_x= w(t,x),\ll{dmkdva}\\
& & w_{x}-2 v(c^2(t)-w^2)^{\frac 1 2}=0.\ll{dmkdvb}\eea
Note that the arbitrary function $c(t) $ can be removed again from the
equation (\ref{dmkdvb}) by rescaling $\frac {w(x,t)}{c(t)} \to w(x,t)  $. We
however keep $c(t)  $ in the explicit form for the later convenience. At
$c(t)=0 $ we should have the deforming function $ w(x,t)=0$,  when the dmKdV
 (\ref{dmkdva}-\ref{dmkdvb}) would reduce to the standard mKdV equation.
 For
establishing the integrability of (\ref{dmkdva}-\ref{dmkdvb}) we derive
the associated  pair of  matrix Lax operators in the form \be
U(\lambda)= U^{mkdv}(\lambda),
 \ V(\lambda)= V^{mkdv}(\lambda)+ 
V^{def}(\lambda), \ll{UVdmkdv}\ee where $U^{mkdv}(\lambda),
V^{mkdv}(\lambda) $
 are the well known Lax pair  for the standard mKdV system \cite{solit1}:
\bea U^{mkdv}(\lambda)&=&i\lambda \sigma^3+i v\sigma^1, \nonumber \\
V^{mkdv}(\lambda)&=& 2 i \lambda (v^2-2 \lambda ) \sigma ^3 -2i
\lambda
 v_x\sigma^2
 + i  (v_{xx}-4 \lambda ^2 v+2 v^3) \sigma^1
,\ll{UVmkdv}\eea while  the additional term \be
V^{def}(\lambda)= \frac i {2 \lambda} (b \sigma^3-w\sigma^2), \ \mbox{where } \
b=(c^2(t)-w^2)^{\frac 1 2} \ll{Vdefmkdv}\ee is responsible for
the  deformation of the mKdV equation.
Here $\sigma^a,\ a=1,2,3 $ are standard Pauli matrices 
\be
 \sigma^1=\left(%
\begin{array}{cc}
  0 & 1 \\
  1 & 0 \\
\end{array}%
\right), \  \sigma^2=\left(%
\begin{array}{cc}
  0 & -i \\
  i & 0 \\
\end{array}%
\right), \ 
 \sigma^3=\left(%
\begin{array}{cc}
  1 & 0 \\
  0 & -1 \\
\end{array}%
\right)
  \ll{pauli}
\ee
and $\sigma^\pm=\frac 1 2 ( \sigma^1\pm i\sigma^2). $
 For confirming the Lax
integrability of the NHD of the mKdV equation
(\ref{dmkdva}-\ref{dmkdvb}) we show that it is derivable from the
flatness condition $U_t-V_x+[U,V]=0 $ of the Lax pair
(\ref{UVdmkdv}). In this process one finds that, the expressions
in all positive powers of $\lambda ^n , n=1,2,3$ vanishes
trivially, while that for $n=0 $ yields the deformed mKdV equation
 (\ref{dmkdva}-\ref{dmkdvb}). The coefficients with the power $n=-1$ gives in turn the
constraint equations \be b_x=-2vw, \ \ w_{x}=2vb \ll{Dgen} \ee
from which eliminating  function $b $ by using $ b=(c^2(t)-w^2)^{\frac 1
2} $ we arrive at the constraint (\ref{dmkdvb}) as required.

 Note that one can also find a single higher order  nonlinear equation by
eliminating further the deforming function $w(x,t) $ from the set of
equations (\ref{dmkdva}-\ref{dmkdvb}).  By introducing  a potential
field $\theta_x=2 v $ we can rewrite Eq. (\ref{dmkdvb}) in the form 
$\theta_x=\frac {  w_{x}} {(c^2-w^2)^{\frac 1 2}},$   giving  an easy 
 solution   $w=c(t)
\sin \theta $,  which can be checked by direct substitution, since $
w_x=c \theta_x \cos \theta $ while $  {(c^2-w^2)^{\frac 1 2}}=c \cos \theta
$.  Inserting these expressions of $w $, $v$  and hence
those of  $v_t,v_{xxx}, v^2v_x $ through $\theta $ in Eq. 
 (\ref{dmkdva})
  one can derive   a new integrable 4-th order
potential mKdV equation in $\theta $ given by
 \be
(\theta_{t}-\theta_{xxx}-\frac 1 2 \theta_{x}^3 )_x= 2c(t)\sin
\theta .\ll{4mkdv}\ee
Interestingly, keeping only the term  $\theta_{xt} $  in the LHS, while
 grouping  the  other terms as 
\be
 f(\theta)=(\theta_{xxx}+\frac 1 4 \theta_{x}^3 )_x
,\ll{pfSG}\ee
 one can rewrite   
Eq. 
(\ref{4mkdv}) in the form of
 a perturbed sine-Gordon (SG)
equation in the light-cone coordinates as
 \be \theta_{xt}= 2c(t)\sin
\theta +f(\theta), \ll{pSG}\ee
 with a time-dependent mass parameter
$m^2=2c(t) $,  perturbed by a  function (\ref{pfSG}). 
 Significantly unlike most of the perturbed SG equation
(\ref {pSG}-\ref {pfSG}), derived here, is integrable allowing exact N-soliton
solution. Thus the integrable higher order dmKdV equation
(\ref{4mkdv}) is equivalent to the NHD of the dmKdV equation
(\ref{dmkdva}-\ref{dmkdvb}), which in turn is equivalent to the
set of equations (\ref{dmkdva}) with  (\ref{Dgen}).

\subsection{Exact soliton solutions}

As is well known any perturbation usually spoils the integrability
of a nonlinear system and hence forbids general  analytic  solutions.
However the dmKdV we have constructed retains its complete  integrability
in spite of the perturbation with nonholonomic constraint.
Therefore
  exploiting this integrability property
we intend to derive  exact N-soliton solutions  for  the    dmKdV
equation (\ref{dmkdva}-\ref{dmkdvb}) or equivalently for the novel
4-th order potential mKdV equation (\ref{4mkdv}),
    through the IST method. The procedure
  follows  that for  the
standard mKdV equation \cite{wadati72} in its initial steps, while in
the final step
 the effect of deformation should be incorporated.
It is intriguing to note  from   (\ref{UVdmkdv}),
 that for the 
dmKdV equation,  only the time Lax operator $V $ is deformed,
 while  the space part $U $
is kept unchanged. This  reveals  an important connection between
the
 deformed time evolution of the Jost solution and the NHD
 of the nonlinear equation.

 Recall  that for extracting the exact soliton solution in
  the IST  the space-Lax operator $U(\la ) $ describing the
scattering of the Jost functions, plays the key role.
 Only at the final stage we need to fix   the time
evolution of the solitons through  the time-dependence of the
spectral data, which   in turn is  determined
 from the asymptotic value of the time-Lax operator
$V(\la )|_{|x| \to \infty}$. Therefore following \cite{wadati72} we
can derive the N-soliton solution for our deformed mKdV equation
as \be v(x)= [ \frac {d^2} {dx^2}\ln det A(x)]^{\frac 1 2},
\ll{Nsol}\ee
where  the matrix function $A(x) $ is expressed
through its elements as
 \be A_{nm}=\sum_{l=1}^N f_{nl}f_{lm}, \
\mbox{where}\ f_{nm}=\delta_{nm}+
 \frac {\beta_n \beta_m}
{\kappa_n+\kappa_m}e^{-(\kappa_n+\kappa_m)x}. \ll{AN}\ee Here
parameters $ \kappa_n, n=1,2,\ldots, N$, denote the
time-independent zeros   of the scattering matrix element:
$a(\la=\la _n )=0 $, along the imaginary axis: $\la_n=i\kappa_n $
and   $ \beta_n(t)=b(\la = \la _n) $ are the time-dependent
spectral data to be determined from $
V(\lambda)=V^{mkdv}(\lambda)+V^{def}(\lambda) $, at $ x \to \pm
\infty$.

 Note that due to the boundary condition (BC) $v \to 0, w \to
0 $ at $ x \to \pm \infty$,
  the asymptotic value of (\ref{UVmkdv}):  $V^{mkdv}(\lambda) \to
-4i\lambda^3 \sigma^3 $ corresponds to the undeformed part,
 while  the BC $b \to c(t) $  at $ x \to \pm \infty$,
determines   the crucial effect of deformation
$V^{def}(\lambda) \to \frac {i} 2 \lambda^{-1} c(t)\sigma^3  $. As a result we
obtain \be \beta_n(t)=\beta_n(0) e^{(8\kappa_n^3t +
 \frac {\tilde c(t)}{\kappa_n})}, \ \tilde c_t(t)= c(t) ,
 \ll{bett}\ee yielding  finally  the $(x,t) $ dependent
exact soliton solution  from (\ref{AN}).

To see  the effect of  deformation  on the dynamics of the
   solitons      more closely
we construct 1-{\it soliton} solution for  the dmKDV
 equation as reduced from  (\ref {Nsol}-\ref{bett})
 at $N=1 $:
\be v(x,t)=\frac {v_0} 2 {\rm sech}\xi,\ \xi= \kappa(x-vt)+\phi, \ v=v_0+v_d
\ll{1s} \ee 
where the  phase  $ \phi$ is an arbitrary  constant, 
 $v_0=4\kappa^2 $ is the usual constant
velocity of the
 mKdV soliton, while
$v_d= \frac {2\tilde c(t)}{v_0 t}$ is the unusual time-dependent
part of the velocity, induced by the  deformation. 
Note that the
time-dependent asymptotic value of  the deformation  acts here
like a forcing term sitting  at the space boundaries, which
  for $c(t)=c_0 t $  with
$c_0>0$ forces the soliton  to accelerate, while with $c_0<0 $
makes it  decelerate (see Fig 1).

\begin{figure}[c]

\includegraphics[width=6.cm,height=6.1 cm]{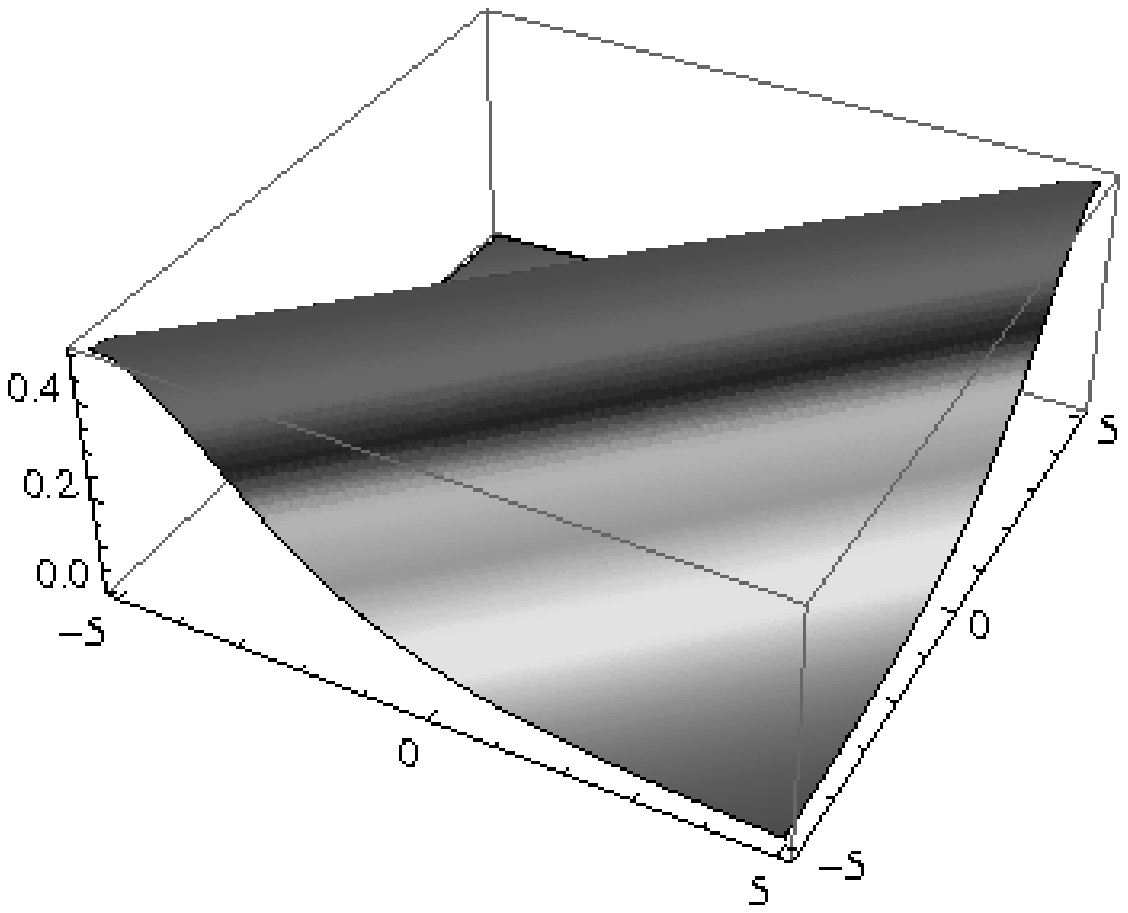}
\ \ \ \ \ \ \ \ \ \ \ \ \ \
 \ \  \includegraphics[width=6.cm,height=6.1 cm]{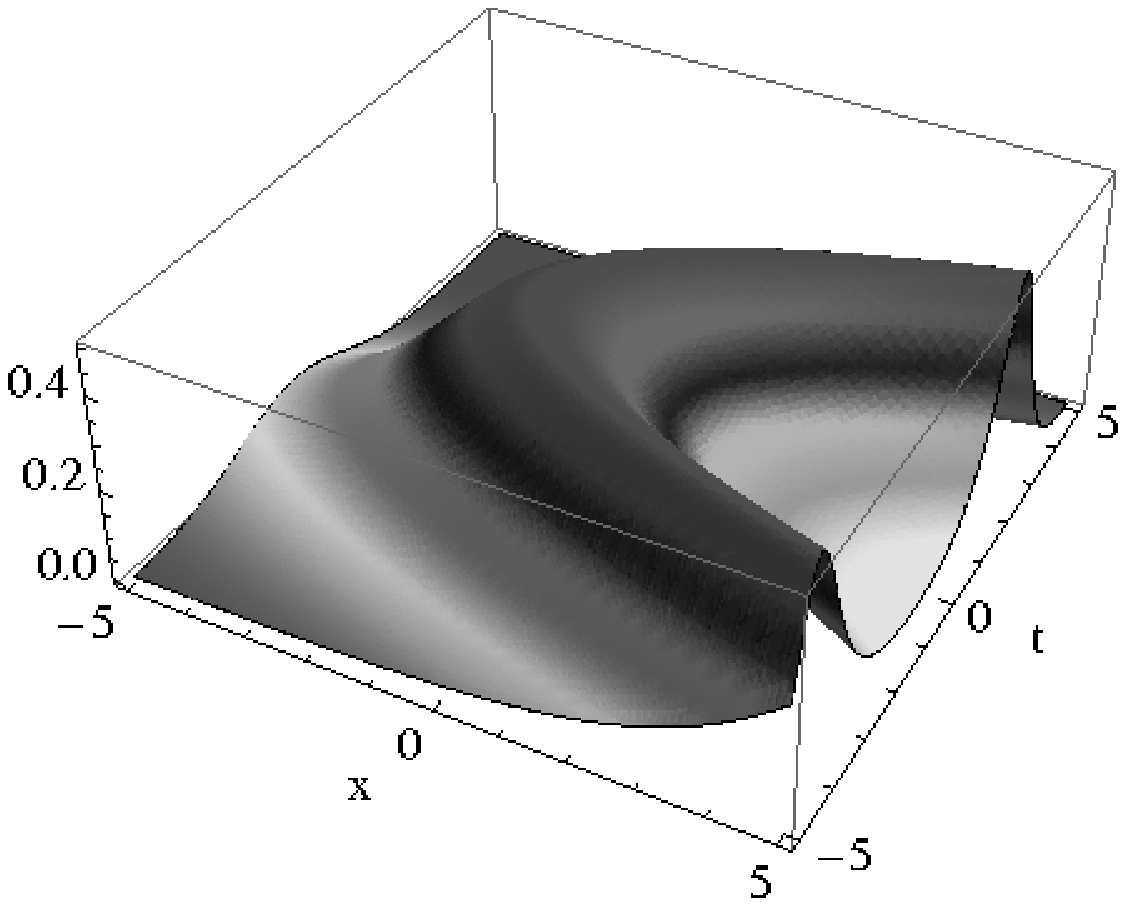}

\caption{  Exact soliton solutions  in the (x,t)-plane:  
 a) for the standard  mKdV  equation with constant  soliton
velocity (indicated by its constant inclination),
 b) for the nonholonomically deformed dmKdV equation  with accelerated
soliton 
(reflected in its  curvature with $c(t)=0.5 t $). }

\end{figure}

It is important to note, that the evolution of the basic field $v(x,t) $ comes here
from two sources:
  $\partial_{t}v=\partial_{t_0} v+\partial_{t_d}v $,  induced by two different
{`times'} $t_0 $ and
$t_d $. $\partial_{t_0} v=\partial_{t}v|_{c(t)=0}$ is the
 evolution due to the unperturbed time $t_0 $, caused by the standard  dispersive and nonlinear terms
in (\ref{dmkdva}), while $\partial_{t_d}v=c(t)\partial_{\tilde c(t)}v $ is
the evolution due to the deformed time $t_d $, caused by the perturbation  
linked as
\be
w=\partial_{t_d}v 
, \  \ b=c(t)+\int v \partial_{t_d}v  dx.
\ll{wb}\ee
Therefore using (\ref{wb}) and $ \partial_{\tilde c(t)} \beta_n(t)
=\frac 1 \kappa_n \beta_n(t) $  we can find from (\ref {Nsol}-\ref{bett})
the N-soliton solution corresponding to  $ w,b$, which for $N=1$ derived
from (\ref{1s}) gives also localized  accelerating  solution 
\be
w(x,t)=c(t)\kappa {\rm sech}\xi {\rm tanh}\xi \ \mbox{and} \
 b(x,t)=c(t)(1-\kappa ^2 {\rm sech}^2\xi  ).
\ll{wb1s}\ee
This shows that the perturbing function, taking itself
the  solitonic form  drives  the field soliton
to have an accelerated motion, while    
 in turn the basic field solution
  self-consistently determines  the solitonic form of the  perturbing function, sustaining
thus the integrability of the system.

\subsection{Generalized symmetries}

It is well known that the Lie symmetry analysis plays an effective role to 
study the integrability properties of nonlinear evolution equations in (1+1)
dimensions such as the 
existence of infinitely many generalised symmetries,  conserved quantities and a recursion 
operator (\cite{ovi}, \cite{olv}, \cite{Blu1}, \cite{fokas}, \cite{Lak}).
 We show here that similar analysis can be performed with
equal success for our deformed mKdV equation
 (\ref{dmkdva}-\ref{dmkdvb}).

Note that the dmKdV equation
is invariant under the scaling or dilatation symmetry
$$(t,x,v,w,c(t))\rightarrow
(s^{-3}t,s^{-1}x,s^{1}v,s^{4}w,s^{4}c(t)), $$ where $s $ is an arbitrary
parameter which suggests that $v$ corresponds to one
derivative with respect to scaling $x$, $w$ corresponds to three
derivatives with respect to  $x$ and $c(t)$ corresponds to four
derivatives with respect to $x$.  We would like to mention that Hereman
 and his collaborators have developed an algorithm to derive generalised symmetries,
 conserved quantities and recursion operators for nonlinear partial 
differential and differential-difference equations \cite{Here1}.
 Hereman's algorithm is based  basically on the concept of weights and ranks. 
The weight $W$ of a variable is defined as the
exponent in the scaling parameter $s$ which multiplies the variable.
 Weights of the dependent variables are non-negative and rational.
An expression is said to be uniform in rank if all its terms have the same rank.
Setting   $W(D)=W(\frac{\partial}{\partial x})
= 1$, we see that $W(v)=1, W(w)=4, W(b)=4, W(c(t))=4$ and
$W(\frac{\partial}{\partial t})=3 $ and hence eqn. (\ref{dmkdva})
is of rank $4$ and   (\ref{dmkdvb})
 is of
rank $5$. This property is called the uniformity in rank.  The
rank of a monomial is defined as the total weight of the monomial,
again in terms of derivatives with respect to $x$.

Now, assume that the deformed mKdV equation
(\ref{dmkdva}-\ref{dmkdvb})
 is invariant under  one parameter nonpoint continuous transformations 
\be t^{*} = t,\ \  x^{*}=x, \ \  v^{*}=v+\epsilon G_{i} + O(\epsilon
^2), \ \ w^{*}=w+\epsilon H_{i} +O(\epsilon^2),\ee where
$$ G_{i} =
G_{i}(v,w,v_{x},v_{xx},v_{xxx},v_{4x},..,w_{x},w_{xx},\cdots)$$
$$H_{i} =
H_{i}(v,w,v_{x},v_{xx},v_{xxx},v_{4x},..,w_{x},w_{xx},\cdots)$$
provided $v(x,t)$ and $w(x,t)$ satisfy equation   (\ref{dmkdva}-
\ref{dmkdvb}).
 Consequently we obtain
the following invariant equations \be
 \frac{D G_{i}}{D t} =
  \frac {D^3 G_{i}}{D x^3}
+6v^2\frac{D G_{i}}{D x}+ 12vv_xG_i
  +\frac {D H_{i}} {D x},
\ll{2.7a}\ee \be \frac{D^2 H_{i}}{D
x^2}=2G_{i}\sqrt{(c(t)^2-w^2)}-2vw \frac{ \frac {D H_{i}} {D
x}}{\sqrt{(c(t)^2-w^2)}},\ll{2.7b}\ee

 where
$$\frac{D}{Dx}=\frac{\partial}{\partial{x}}+v_x\frac{\partial}{\partial{v}}+
w_x\frac{\partial}{\partial{w}}+v_{xx}
\frac{\partial}{\partial{v_x}}+v_{xt}\frac{\partial}{\partial{v_t}}
+w_{xx}\frac{\partial}{\partial{w_x}}+
w_{xt}\frac{\partial}{\partial{w_t}}+\cdots$$
$$\frac{D}{Dt}=\frac{\partial}{\partial{t}}+v_t\frac{\partial}{\partial{v}}+
w_t\frac{\partial}{\partial{w}}+v_{xt}
\frac{\partial}{\partial{v_x}}+v_{tt}\frac{\partial}{\partial{v_t}}
+w_{xt}\frac{\partial}{\partial{w_x}}+
w_{tt}\frac{\partial}{\partial{w_t}}+\cdots.$$
 From equations (\ref{2.7a}-\ref{2.7b}), we see
that

\be G_{1}=v_{x},\ \ \  H_{1}=w ,\ll{2.8} \ee is a trivial
generalized symmetry with rank $(2,4)$.  This suggests that the
next generalized symmetry $G_{2}$ and $H_{2}$ of dmKdV must have
rank (4,6).  With this in mind we first form monomials in $v$ and
$w$ of rank $(4,6)$, Thus the most general form of $ G_{2}$ and
$H_{2}$ will be \be G_{2}= a_{1}v^2v_{x} +a_{2} v_{3x}+a_{3}w
, \ \ H_{2}= b_{1}v^2w +b_{2}w_{2x}, \ee where
$a_{i},b_i, i=1,2,3$ are arbitrary constants to be determined.  We now substitute
$G_{2}$ and $H_{2}$ in the invariant equation
(\ref{2.7a}-\ref{2.7b}),with $i=2$ and using (\ref{dmkdva}-\ref{dmkdvb})
we find that the consistency condition holds only for the following parametric
restrictions:
$$a_{1}=6,a_{2}=1,a_{3}=0,b_{1}=6,b_{2}=1$$ and so the generalized symmetry with rank $(4,6)$ becomes \be G_{2}=
6v^2v_{x} +v_{3x}, \ \ \ H_{2}= 6v^2w +w_{2x}. \ee
Proceeding as above, for $i = 3$, we find that the invariant
equations (\ref{2.7a}-\ref{2.7b}) satisfy only if \be
G_{3}=v_{5x}+10v_{x}^3+40vv_xv_{2x}+10v^2v_{3x}+30v^4v_{x}, \ \ \
\ \ \ee \be
H_{3}=w_{4x}+10v_{x}^2w+20vv_{2x}w+20vv_xw_{x}+10v^2w_{2x}+30v^4w ,\ee
which is a next nontrivial generalized symmetries with rank( 6,8).
In a similar manner we can derive infinitely many generalized symmetries
$\{(G_{4},H_{4}), (G_{5},H_{5}),\cdots\}$ for (\ref{dmkdva}-\ref{dmkdvb}) with rank
$\{(8,10),(10,12),\cdots\}$.  We have also checked
that the commutator$$[G_{i},G_{i+1}]=0,[H_{i},H_{i+1}]=0,  \ \
\forall \ i,
$$ indicating that  dmKdV eqn. (\ref{dmkdva}-\ref{dmkdvb}) admits an infinitely many
generalized symmetries which commute.

\subsection{Recursion operator}
In this section we derive a Recursion operator, an important property of
the integrable systems, for our  deformed mKdV
eqn. (\ref{dmkdva}-\ref{dmkdvb}), which  is usually possible to obtain
only
for the unperturbed integrable systems. An operator valued function
$\mathcal{R}$ is said to be a recursion operator of a scalar
nonlinear partial differential equation with two independent
variables if it satisfies
$$\tilde{G}=\mathcal{R} G,$$
where $\tilde {G}$ and $G$ are successive generalized symmetries.
For (\ref{dmkdva}-\ref{dmkdvb})
the above equation can be written as\\
\be \left[%
\begin{array}{c}
 G_{k+1} \\
 H_{k+1} \\
\end{array}%
\right] =\mathcal{R} \left[%
\begin{array}{c}
  G_{k} \\
  H_{k} \\
\end{array}%
\right] = \left(%
\begin{array}{cc}
  R_;{11} & R_{12} \\
  R_{21} & R_{22} \\
\end{array}%
\right)\ \left[%
\begin{array}{c}
 G_{k} \\
 H_{k} \\
\end{array}%
\right], \ll{2.9}\ee where $(G_{k},H_{k})$ and $(G_{k+1},H_{k+1})$
are successive generalized symmetries and $R_{ij},i,j=1,2$ are
 functions of dependent variable and their differential and integral operators. 
 The construction of the recursion
operator $\mathcal{R}$ for the dmKdV equation is as follows: For
$k=2$ equation (\ref{2.9}) becomes
\be \left[%
\begin{array}{c}
 G_{3} \\
 H_{3} \\
\end{array}%
\right] = \left(%
\begin{array}{cc}
  R_{11} & R_{12} \\
  R_{21} & R_{22} \\
\end{array}%
\right)\ \left[%
\begin{array}{c}
 G_{2} \\
 H_{2} \\
 \end{array}%
  \right]
  \ll{2.9a} \ee where $(G_{2},H_{2})$ and $(G_{3},H_{3})$ are
the generalized symmetries of ranks $(4,6)$and $(6,8)$
respectively. The ranks of $R_{11},R_{12},R_{21}$ and $R_{22}$ can be
determined from the following relations
 \be rank G_{3}=rank R_{11}+rank
G_{2}=rank R_{12}+rank H_{2}, \ll{2.10a}\ee
 \be rank H_{3}=rank
R_{21}+rank G_{2}=rank R_{22}+rank H_{2},\ll{2.10b}\ee Equations
(\ref{2.10a}-\ref{2.10b}) show that the ranks of
$R_{11},R_{12},R_{21},R_{22}$ respectively are $2,0,4, 2$ and so
we consider the entries of $\mathcal{R}$ written in terms of
differential and integral operators of the dependent variables
having the form \bea R_{11}&=& c_{0}\partial^{2}+c_{1}v^2+c_{2}v_{x}\partial^{-1}v
,\ R_{12} = 0,  \ R_{22}=g_{0}\partial^{2}+g_{1}v^2+g_{2}v\partial^{-1}v_x,\nonumber \\
R_{21}&=&
f_{0}w\partial^{-1}v+f_{1}v\partial^{-1}w+f_{2}v^4+f_{3}w
+f_{4}v_{4x}\partial^{-1}+f_{5}v^2\partial^{2},
 \ll{Rs}
\eea where$
c_{i},g_{i},i=0,1,2$ and  $f_{j},j=0,1,2,3,4,5$ are constants to be determined.  Substituting the above
in  (\ref{2.9a}) we find that it is satisfied identically only if
$$c_{0}=1, c_{1}=4, c_{2}=4,
f_{0}=4,f_{1}=4,f_{2}=f_{3}=f_{4}=f_{5}=0,
g_{0}=1,g_{1}=4,g_{2}=-4$$ and therefore
 the recursion operator for the
dmKdV equation becomes
\be \mathcal{R}=\left(%
\begin{array}{cc}
  \partial^2+4v^2+4v_{x}\partial^{-1}v & 0 \\
  4w\partial^{-1}v+4v\partial^{-1}w & \partial^2+4v^2-4v\partial^{-1}v_{x} \\
\end{array}%
\right). \ee

\subsection{Conserved quantities}
A local conservation law of a partial differential equation with two
independent variables $(x,t)$  is defined by \be
\frac{\partial}{\partial t}\rho + \frac{\partial}{\partial x}J = 0
\ll{cont} \ee which is satisfied on all solutions. The function $\rho(x,t)$ is
usually called local conserved density and $J(x,t)$ is the
associated flux also known as current density. We show that the dmKdV (\ref{dmkdva}-\ref{dmkdvb}) or
eqn.(3) and (8) admit infinitely many polynomial conserved quantities. From
 (\ref{dmkdva}), we find directly  that

\be \rho[1]=v,   \ \ \ \    J[1]= -(v_{xx}+2v^3+\int dx w) \ee is a trivial
conserved quantity with rank $(1,3)$. Recall that (\ref{dmkdva}) and
(8) is invariant under scaling symmetry \ $(t,x,v,w,b,c)\rightarrow
(s^{-3}t,s^{-1}x,s^{1}v,s^{4}w,s^{4}b,s^{4}c), \ $ where $s $ is an
arbitrary parameter.
 To derive a conserved quantity with rank
$(2,4)$,  as before, we form monomials of $v(x,t)$ and $w(x,t)$ which
gives the list $\mathcal{L}_{1} = \{ v^2 \}$. Thus the conserved
density of rank 2 will be
$\rho(x,t) = v^2. $ As a result we
obtain
 \be \rho[2]=v^2,  \ \
J[2]=-2vv_{xx}-3v^4+v_{x}^2+b. \ll{c2}\ee

 Proceeding as above we find the next two
conserved quantities as \be \rho[3] = v^4-v_{x}^2,
J[3]=-4v^6-4v^3v_{xx}+2v_{x}v_{xxx}-v_{xx}^2+12v^2v_{x}^2+2v^2b, \ll{c3} \ee
\bea  \rho[4]&=&v^6-5v^2v_{x}^2+\frac{1}{2} v_{xx}^2 ,\\
J[4]&=&
\frac{-9}{2}v^{8}-6v^{5}v_{xx}+10v^2v_{x}v_{xxx}-v_{xx}v_{4x}
+\frac{1}{2}v_{xxx}^2
-8v^2v_{xx}^2 \nonumber \\ &+&45v^4v_{x}^2
-10vv_{x}^2v_{xx}-\frac{1}{2}v_{x}^4+3v^4b -v_{x}^2b-2vv_{x}b_{x}.
\eea
with ranks (4,6) and (6,8) respectively. 
In a similar manner we can derive  an infinitely many conserved quantities 
$(\rho(x,t),J(x,t))$ for dmKdV with ranks
$\{(8,10),(10,12) \ldots  \}$ which involve lengthy expressions and so the
details are omitted here.

It is important to notice that the integrals of motion describing
infinite number of integrated conserved quantities $c_n, \ n=1,2,\ldots $,
which should be  commutative as a necessary criterion of Liouville
integrability  can be  given by $c_n=\int \rho[n] dx $. It is clearly seen
from the continuity equation (\ref{cont}) that due to vanishing of the
fields  along with their derivatives at the space-infinities we naturally
obtain $\partial_t c_n=0 $.
Therefore from the expressions of  $\rho[n], n=1,2,\ldots $, derived above
we get the conserved quantities for the dmKdV equation  as   \be
c_1= \int v dx, \ c_2= \int  v^2 dx, \ c_3= \int ( v^4-v_{x}^2)dx
 \ll{cn}\ee
with $c_3=H_{mkdv} $ being the Hamiltonian of the mKdV equation.
We see therefore that the conserved quantities including the Hamiltonian
remains the same for both the deformed and undeformed mKdV systems, though
the corresponding equations for the dmKdV with additional perturbing
function and nonholonomic constraint on it are surely   different.
Note also that this effect of deformation changes    the structure of
the local current densities $J[n],n=1,2,\ldots $, which contain the deforming
functions
$w,b $, but  not  the  densities $\rho[n] $, which generate the
conserved quantities.
\section{Nonholonomic deformation of the KdV equation and its
integrability aspects} Nonholonomic constraints on field models
have received
 increasing attention over
recent years
, while  a  significant breakthrough is made recently by
discovering an integrable  nonholonomic deformation for the
   KdV equation
 \cite{karasu,kuper08,kundu082,Yao08}.
It has been established that    this deformed KdV equation
 like the undeformed standard KdV  admits Lax pair,
 N-soliton solutions and a two-fold integrable hierarchy \cite{kundu082}.
  The deformed KdV can be considered also as a source equation \cite{Yao08},
 which is
however different from and simpler than the well known source KdV
equation
 \cite{melnikov}. The
novelty of this  source is that it can be
 deformed recursively by going to the next order in its
 integrable hierarchy with higher order deformations \cite{kundu082,kundu081}.

However  many other  fundamental and important properties of the
integrable systems, such as the existence of an infinite number of  generalized
symmetries,  conserved quantities and a recursion
operator, which have been well
established for the standard KdV  equation have not yet
been studied for their nonholonomically deformed extension, investigation of
which is
therefore  our aim here.

\subsection{Generalized symmetries}
 The deformed KdV equation  (\ref{dkdva}-\ref{dkdvb}) is obviously
invariant under the dilatation symmetry \be (t,x,u,g,c(t))\rightarrow
(s^{-3}t,s^{-1}x,s^2u,s^4 g,s^4 c(t)),\ll{4.1} \ee where $s $ is an
arbitrary parameter which suggests that $u$ corresponds to two
derivatives, while both  $g$ and $c(t)$ correspond to
four derivatives, with respect to $x$. Setting
$W(D)=W(\frac{\partial}{\partial x}) = 1$, therefore one gets   $W(u)=2, W(g)=4,
W(c(t))=4$ and $W(\frac{\partial}{\partial t})=3 $ and hence equations 
(\ref{dkdva}-\ref{dkdvb})
 is of rank $(5,7)$. This property is called the uniformity in rank.
Now, assume that the deformed KdV equation
(\ref{dkdva}-\ref{dkdvb})
 is invariant under  one parameter continuous nonpoint
transformations \be u^{*}=u+\epsilon G_{i} +
O(\epsilon ^2),\ll{4.2a}\ee \be g^{*}=g+\epsilon H_{i}
+O(\epsilon^2),\ll{4.2b}\ee where
$$ G_{i} =
G_{i}(u,g,u_{x},u_{xx},u_{xxx},u_{4x},..,g_{x},g_{xx},\cdots)$$
$$H_{i} =
H_{i}(u,g,u_{x},u_{xx},u_{xxx},u_{4x},..,g_{x},g_{xx},\cdots)$$
provided $u(x,t)$ and $g(x,t)$ satisfy equations   (\ref{dkdva}-\ref{dkdvb}).
 Consequently we obtain
the following invariant equations \be
 \frac{D G_{i}}{D t} =
  \frac {D^3 G_{i}}{D x^3}
+6u\frac{D G_{i}}{D x} +6u_{x} G_{i} +\frac {D H_{i}} {D x},
\ll{4.3a}\ee \be \frac{D^3 H_{i}}{D x^3}+ 4u \frac{D H_{i}}{D x}+4
G_{i}g_x +2\frac{D G_{i}} {D x} (g+ c(t))
+2u_xH_{i}=0.\ll{4.3b}\ee
 From (\ref{4.3a}-\ref{4.3b})  we derive  the following generalized
 symmetries
 \be G_{1}=u_{x}, H_{1}=g_{x} \ \ G_{2}= 6uu_{x} +u_{3x}, H_{2}=
6ug_{x} +g_{3x},  \ee
which are  trivial ones with ranks $(3,5)$ and $(5,7)$, respectively. Proceeding as before
we find that
(\ref{dkdva}- \ref{dkdvb}) admits an infinitely many generalized
symmetries, where  the first two nontrivial generalised symmetries
are 
$$G_{3}=30u^2u_{x}+20u_{x}u_{xx}+10uu_{3x}+u_{5x}, $$
$$H_{3}=30u^2g_{x}+10u_{x}g_{xx}+10u_{2x}g_{x}+10ug_{3x}+g_{5x},$$
and $$G_{4}=
140u^3u_{x}+70u_{x}^3+280uu_{x}u_{2x}+70u^2u_{3x}+70u_{2x}u_{3x}+42u_{x}u_{4x}+14uu_{5x}+u_{7x}$$
$$ H_{4}=
140u^3g_{x}+70u_{x}^2g_{x}+140uu_{x}g_{2x}+140uu_{2x}g_{x}+70u^2g_{3x}$$
$$\hspace*{1.5cm}+42u_{2x}g_{3x}+28u_{3x}g_{2x}+28u_{x}g_{4x}+14u_{4x}g_{x}+14ug_{5x}+g_{7x}$$
with ranks $(7,9)$ and $(9,11)$, respectively. Note that
 the remaining higher order generalised symmetries involve lengthy expressions and hence not
presented  here.  We find also
that the commutator relations
$$[G_{i},G_{i+1}]=0,\ \ \ [H_{i},H_{i+1}]=0,\forall i $$ hold.
It is straightforward again to check that the derived generalised symmetries satisfy $$ \left[%
\begin{array}{c}
 G_{i+1} \\
 H_{i+1} \\
\end{array}%
\right] =\mathcal{R} \left[%
\begin{array}{c}
  G_{i} \\
  H_{i} \\
\end{array}%
\right] \forall i,$$ with the recursion operator
\be \mathcal{R}=\left(%
\begin{array}{cc}
  \partial^2+4u+2u_{x}\partial^{-1} & 0 \\
  2g_{x}\partial^{-1}+2\partial^{-1}g_{x} & \partial^2+4u-2\partial^{-1}u_{x} \\
\end{array}%
\right). \ee
\subsection{Conserved quantities}
 Recall that the dKdV equation  (\ref{dkdva}-\ref{dkdvb}) is invariant
under the dilatation symmetry $ (t,x,u,g)\rightarrow
(s^{-3}t,s^{-1}x,s^2u,s^4 g). $  We can show again 
that the dKdV equation admits an infinitely many conserved
quantities. From (\ref{dkdva}-\ref{dkdvb}), we find that

\be \rho[1]=u,   \ \   J[1]=-(u_{xx}+3u^2+g) \ee correspond to a trivial
conserved quantity with rank $(2,4)$. To derive the next conserved
quantity with rank $(4,6)$   we form monomials of
$u$ and $g$ as before, which give the list $\mathcal{L}_{1} = \{
u^2,u_{xx} \}$. Thus the most general form of the conserved
density of rank 4 would be
\ $\rho(x,t) = c_{1} u^2 + c_{2} u_{xx} ,$ \ with $c_{1}$,  $c_{2} $ are constants. It is straightforward 
to check that the next order 
conserved density and associated flux \be
\rho[2]=u^{2}, \ J[2]=(-4u^3-2uu_{xx}+u_{x}^2+2ug+g_{xx}+2uc(t)) \ee
satisfy (\ref{cont}). In a similar manner we obtain in the next higher  order
$$ \rho[3]=(4u^{3}-2u_{x}^2), \ J[3]=(-18u^4+24uu_{x}^2-12u^2u_{xx}-2u_{xx}^2
+4u_{x}u_{3x}$$\be
-g_{4x}+4u^2g+4u^2c(t)-2u_{xx}c(t)-2u_{xx}g-6u_{x}g_{x}). \ee We
wish to add that the next higher order conserved quantities
involve lengthy expressions and so we refrain from presenting them.

We can  obviously construct the commuting family of integrated  conserved quantities
as above:
 $c_n=\int \rho[n] dx $ with $\partial_t c_n=0 $.
Therefore from the corresponding  $\rho[n], n=1,2,\ldots $,
we get the conserved quantities for the dKdV equation  as   \be
c_1= \int u \ dx, \ c_2= \int u^2 \ dx , \ c_3= \int  (4u^3-2u_{x}^2)dx, ...
 \ll{cnn}\ee
with $c_3=H_{kdv} $ being the Hamiltonian of the KdV equation.
Therefore we see again  that for the deformed KdV system, the
  conserved quantities
remain the same as in the undeformed case, though the corresponding
  equation gets deformed
due to the nonholonomic constraint.

\section{ Generalization to the nonholonomic deformation of the  AKNS
system } In sect. 2 we have discovered  a new
integrable
 nonholonomic deformation of the mKdV system, extending the dKdV. Here
 we intend to show that   these
deformed systems  can be unified and generalized  further to an integrable
nonholonomic deformation of the AKNS \cite{solit1} system: \be q_t-q_{xxx}-6
(qr)q_x=g_x, \ll{3aknssq}\ee \be r_t-r_{xxx}-6 (qr)r_x=f_x.
\ll{3aknssr}\ee For finding the integrable nonholonomic constraints on the
deforming functions $g, f $ we introduce the deforming matrix $
G^{(1)}=g\sigma^++f\sigma^-+b\sigma^3$ and  similarly  for $
G^{(2)}$, and denote the AKNS fields
 in the matrix
form $U^{(0)}=q\sigma^++r\sigma^- $. Integrability condition, i.e. the
flatness condition of the associated deformed Lax pair (with  the additional
 deformation
 $V^{(def)}_2(\lambda)
=\frac i 2 (\lambda ^{-1}G^{(1)}+\lambda ^{-2}G^{(2))}$),
therefore
leads to the   nonholonomic  constraint: \be
G^{(1)}_x=i[U^{(0)},G^{(1)}]+i[\sigma_3, G^{(2)}],\
G^{(2)}_x=i[U^{(0)},G^{(2)}].
 \ll{EE} \ee
 Higher order  integrable deformations can be generated
  recursively by adding to $V^{(def)}_2(\lambda)$ more  terms like
 $\frac i 2 \lambda ^{-j}G^{(j)}, j =3, \ldots n$ with arbitrary $n$. It
 would
result to  a new integrable hierarchy of nonholonomic
deformations of the AKNS (dAKNS) system, given explicitly through the
higher order constraints
  \bea
 G^{(1)}_x&=&i[U^{(0)},G^{(1)}]+i[\sigma_3, G^{(2)}], \ \ldots \
 ,
\nonumber \\G^{(n-1)}_x&=&i[U^{(0)},G^{(n-1)}]+i[\sigma_3,
G^{(n)}], \nonumber \\ G^{(n)}_x&=&i[U^{(0)},G^{(n)}].
 \ll{EEn} \eea
  For $n=1$ this hierarchy clearly reduces to
\be   G^{(1)}_x=i[U^{(0)},G^{(1)}],
 \ll{E} \ee
while
    for
 $
 n=2$ gives  (\ref{EE}).

The integrability of  the dAKNS system
(\ref{3aknssq}-\ref{EE}) is guaranteed from its associated
matrix Lax pair \bea U(\lambda)&=&i\lambda \sigma^3+i U^{(0)}, \
 U^{(0)}=q\sigma^++r\sigma^-, \nonumber \\
 V(\lambda)&=& V^{AKNS}(\lambda)+V^{def}(\lambda),  \ll{UVakns}\eea
where \bea
V^{AKNS}(\lambda)&=& iU^{(0)}_{xx}-4i\lambda^3\sigma ^3+2 \sigma ^3
\lambda (
-U^{(0)}_x+i(U^{(0)})^2)   -4i  U^{(0)}\lambda^2  \nonumber \\
&+&2i
(U^{(0)})^{3} -[ U^{(0)}, U^{(0)}_x], \nonumber \\ 
  V^{def}(\lambda)&=& \frac 1 2(
\lambda^{-1}G^{(1)}+ \lambda^{-2}G^{(2)}). \ll{UV1akns}\eea
  We can check that
 the flatness condition  $U_t-V_x+[U,V]=0 $ of this Lax pair
yields  the dAKNS system  (\ref{3aknssq}-\ref {EE}).
Note that we can generate a novel two-fold integrable hierarchy 
 for  this deformed AKNS system.
Firstly,
by keeping the perturbed equations
(\ref{3aknssq}-\ref{3aknssr}) the same, but by
 increasing the order of the 
 differential
constraint recursively as (\ref{EEn}), we obtain a new integrable
hierarchy for the dAKNS. Secondlly,
  by keeping the constraint fixed  to its lowest level, i.e. as
(\ref{E}) or (\ref{EE}), one can increase also the order of the AKNS
equation with higher
dispersions in the lefthand side of (\ref{3aknssq}-\ref{3aknssr})
in the standard way, generating another 
integrable hierarchy.

This general deformed AKNS system along with its two-fold
integrable hierarchy
 proposed here  can yield as particular cases  both the
 dmKdV and the dKdV equations considered above.
One  can check that the dmKdV equation   proposed here can be
obtained directly from the dAKNS
  through  the reduction  $ q=r=v$, which
degenerates   (\ref{3aknssq}-\ref{3aknssr}) to the
same equation (\ref{dmkdva}), while the constraint (\ref{dmkdvb})
is derived from (\ref{E}) with 
\be G^{(1)}=i (c^2(t)-w^2)^{\frac 1
2} \sigma^3+iw\sigma^2 , \ \   G^{(2)}=0. \ll{G1} \ee
Similarly for deriving the dKdV from the general case of
 dAKNS system we have to consider the particular
reduction
 $ q=u, r=1$ , which obviously  makes (\ref{3aknssr}) trivial
while reduces (\ref{3aknssq}) to the deformed KdV (\ref{dkdva}).
At the same time
 the constraint  (\ref{dkdvb})
 can be derived by solving  (\ref{EE}) as
\bea
G^{(1)}&=&i(g+c) \sigma^3- g_x \sigma^+\nonumber \\
G^{(2)}&=& \frac { g_x} 2 \sigma^3+ i(g+c) \sigma^-+ e\sigma^+ , \
e_x=iu g_x, \ll{G12} \eea where $u(x,t) $ is the KdV field and $
(x,t) $ is the deforming function with  $c(t) $ as   its asymptotic value
$\lim_{|x| \to \infty} g(x,t)=c(t) $.

Reductions of the deformed AKNS system to other important
integrable deformations, e.g. NHD of the nonlinear Schr\"odinger
and sine-Gordon equations have been considered in  \cite{kundu081}.
\section{Concluding remarks}

We have extended the concept of
 nonholonomic deformation from the KdV equation to the mKdV equation and
constructed for this novel deformed system all essential and important
structures of an integrable system,
 like
 a matrix  Lax pair, exact N-soliton solution through IST,
 integrable hierarchies etc.  Interestingly. the integrable  deformed
mKdV, that we have discovered  here,  shows
even a richer picture than that of  the  usual  integrability.

 In particular, it allows accelerating exact solitons
for the basic field as well as for the perturbing function.
The perturbing function, similar to the deformed KdV case \cite{kundu082},
takes a consistent solitonic form and
forces the field soliton  through
its asymptotic value $c(t) $ at space infinities,
to move with an acceleration (or deceleration). Therefore
such a self consistent {\it solitonic} perturbation,
 which  preserves the integrability,
could be significantly important  in  laboratory  experiments with mKdV solitons,
 where the  usual loss of energy inevitable in a realistic  system  could be
compensated for by the driving force given  from the boundaries as found
here.

Moreover, the deformed mKdV system shows  a novel two-fold integrable
hierarchy, as  found  also for the deformed KdV \cite{kundu082}. The first one
corresponds to the standard mKdV hierarchy with higher dispersions but 
perturbed with a deforming function subjected to a given   nonholonomic constraint.
 The second one is a new hierarchy
of equations, where  the same mKdV equation is perturbed now by a function with
increasingly higher order of   nonholonomic constraint.

In parallel with the established procedure for finding the symmetries
of the integrable  equations, we have constructed
 for the first time an infinite  number of 
generalized symmetries, conserved quantities and a recursion operator for
both the deformed    KdV and the deformed mKdV equations. It reveals a remarkable fact
that, though the deforming functions  are contained in  local current
densities,
   the
local densities do not have any dependence on them. Hence 
 the integrated  conserved quantities are also not influenced by the
deformation and   remain the
same as in the unperturbed case.

Finally we have  unified and generalized the deformed KdV and mKdV  equations to 
the deformed AKNS
system, with explicit construction of its Lax pair and a two-fold integrable
hierarchy. Particular reductions of this new deformed integrable AKNS system
yield the deformed KdV as well as  the deformed mKdV system presented here.

The study of the deformed   Painlev\'e class of equations obtained
as   reductions of the dKdV and dmKdV  equations,  a potentially important
project, we wish to   pursue further.

\end{document}